\documentclass[a4paper]{article}

\usepackage{INTERSPEECH2021}
\usepackage{amsmath,graphicx,amssymb}
\usepackage{xspace}
\usepackage{microtype, subfigure, verbatim}
\usepackage{hyperref}
\usepackage{url}
\usepackage{booktabs, multicol, multirow, array} %
\usepackage{arydshln, xcolor}
\usepackage{boldline,pifont}

\def\eg{\textit{e.g.}\xspace}

\def\fsd{\mbox{FSDKaggle2019}\xspace}
\def\magna{\mbox{MagnaTagATune}\xspace}
\newcommand{\cmark}{\ding{51}\xspace}
\newcommand{\xmark}{\ding{55}\xspace}

\title{Do sound event representations generalize to other audio tasks? \\A case study in audio transfer learning}

\name{Anurag Kumar*$^{\dagger}$, Yun Wang*\thanks{* Equal contribution.}$^{\ddagger}$, Vamsi Krishna Ithapu$^{\dagger}$, Christian Fuegen$^{\ddagger}$}
\address{$^{\dagger}$Facebook Reality Labs Research, $^{\ddagger}$ Facebook Applied AI Research}

\email{\{anuragkr,yunwang,ithapu,fuegen\}@fb.com}

\begin{document}
\maketitle

\begin{abstract}
Transfer learning is critical for efficient information transfer across multiple related learning problems. A simple, yet effective transfer learning approach utilizes deep neural networks trained on a large-scale task for feature extraction. 
Such representations are then used to learn related downstream tasks. 
In this paper, we investigate transfer learning capacity of audio representations obtained from neural networks trained on a large-scale sound event detection dataset. We build and evaluate these representations across a wide range of other audio tasks, via a simple linear classifier transfer mechanism. 
We show that such simple linear transfer is already powerful enough to achieve high performance on the downstream tasks. 
We also provide insights into the attributes of sound event representations that enable such efficient information transfer. 
\end{abstract}

\section{Introduction}
\label{sec:intro}

Building task-agnostic representations that can generalize across multiple learning problems has been critical in advancing and applying machine learning techniques in a variety of domains. However, often the design of neural networks are driven by low-level tasks in a given problem domain. For instance, a variety of robust deep networks for low-level tasks like visual object recognition, co-segmentation, etc., have been designed and thoroughly evaluated. While these carefully designed networks are very successful on such low-level tasks, the need for frameworks and algorithmic procedures that combine and {\it transfer} information across multiple low-level tasks to help tackle higher-level tasks (like multi-sensory scene parsing, activity understanding, etc.), remains a challenging problem. Moreover, this notion of combining or sharing knowledge is helpful for training systems under limited and noisy-labeled data as well \cite{wang2020generalizing, lee2018cleannet}. 

Transfer learning is possibly the best suited framework for building such shareable representations, and has been studied comprehensively in the domains of computer vision and natural language processing~\cite{weiss2016survey, tan2018survey, zhai2019large}. Taking vision as an example, transfer learning has been applied to a wide range of problems like scene understanding and action summarizing~\cite{csurka2017domain, zamir2018taskonomy, sargano2017human}, few shot learning and noisy label learning~\cite{wang2020generalizing, lee2018cleannet}, to mention a few. Systems trained on low-level vision tasks such as object detection and classification serve as the source task from which the knowledge is transferred. This is mainly because of the availability of large-scale annotated datasets for such tasks.

In acoustics, specifically, in audio machine learning, transfer learning has been studied to a relatively lesser extent. One possible reason is the less obvious choice of low-level source task. Nevertheless, transfer learning has gained traction in recent times in the audio machine learning. It's been studied in isolated contexts such as sound event detection (SED) \cite{kumar2018knowledge, arora2017study, kong2019panns}, music tagging~\cite{lee2017multi} and emotion recognition~\cite{latif2018transfer}. Nevertheless, we do not yet understand the nuances of generalized representations that capture structural similarity between source and target audio tasks. While prior works have employed transfer learning for different audio tasks, knowledge transfer from a \emph{single low-level audio task } to a variety of other audio tasks has not been studied comprehensively. This forms the key motivation of this paper. Clearly, the capability of shareable representations may depend entirely upon the choice of the tasks used for evaluation. We hypothesize that SED representations to have substantial capabilities to generalize to other related audio task. We choose SED as the source task for two reasons: \emph{first}, sound event datasets are among the largest available audio datasets, thereby providing a large enough database for learning ``robust'' representations; \emph{second}, learning sound events implicitly entails learning low-level acoustic phenomena, which, in principle, amounts to capturing a significant amount of information in an audio snippet. We refer to the SED as the \emph{source} task and explore their generalization power to other audio\emph{target} tasks. Besides, benchmarking capabilities of SED representations for audio transfer learning, we aim to provide interesting insights into the target tasks and the relationship between the target tasks and the source task. 

Keeping the above motivations in mind, we standardized the transfer learning process in the following way. We train neural networks for a large scale SED task and transfer the representations obtained from these networks for any given audio to the target tasks. To reduce the bias in the design of the SED model itself, we train and analyze results through two separate networks. We constrain ourselves to training \emph{linear classifiers} for each target task using the representations obtained from the SED networks. Linear classifiers allow a simple and grounded way to evaluate the efficacy of these audio representations for knowledge transfer.
Even using a simple non-linear mapping for transfer limits us from disentangling the power of sound event representations vs. the power of non-linear transfer itself. Finally, Finally, we consider a variety of target tasks to help to better understand the effectiveness as well as the limitations of these audio representations obtained from SED models. 

In Section~\ref{sec:source-task} we introduce the networks used for SED, and Section~\ref{sec:target-tasks} discusses the target tasks. We evaluate the transfer of event representations in Section~\ref{sec:exp}, and provide insights with some visualizations in Section~\ref{sec:visualization}. Section~\ref{sec:conclusion} concludes the paper.


\vspace{-0.1in}
\section{Source Task \& Audio Representations}
\vspace{-0.1in}
\label{sec:source-task}

As said above, the source task is sound event detection, and representations are obtained from two state-of-the-art deep networks trained on the AudioSet~\cite{gemmeke2017audio} corpus, which contains $2$~million training recordings of $527$ types of sound events. The two models, TALNet~\cite{wang2019comparison} and WEANet-SUSTAIN~\cite{kumar2020sequential}, are briefly summarized below. 

\subsection{TALNet}
\label{sec:talnet}

TALNet~\cite{wang2019comparison} is a deep convolutional recurrent network for SED. The network takes logmel spectrograms as inputs; the logmel spectrograms have $40$~frames per second and $64$~frequency bins. 
The input features are passed through $10$~convolutional layers, $5$~pooling layers, and one bidirectional GRU layer. The output of the GRU has $10$~frames per second, each being a $1,024$-dimensional vector. 
These vectors are further processed by a fully connected layer to calculate the probability of each type of sound event at each frame, and these probabilities are aggregated over an entire recording using a linear softmax pooling function to yield global event probabilities.
We extract the $1,024$-dimensional output of GRU layer (averaged over time) as the learned transferable representation for any given input. Before using these representations to train linear classifiers, we first normalize them to have zero mean and unit variance across training data, then normalize each vector to have unit $l_2$-norm.

\vspace{-0.01in}
\subsection{WEANet-SUSTAIN}
\label{sec:WEANET}

The second network we use is WEANet-SUSTAIN~\cite{kumar2020sequential}. This network also takes $64$-dimensional logmel spectrograms as input, but the frame rate is $100$~frames per second. The network is a fully convolutional neural network with a class-specific attention layer.
The input is first processed by $4$ blocks of layers (B1 to B4); each block consists of $2$ convolutional layers followed by max pooling. These blocks are followed by $4$ more blocks (B5 to B8) of only convolutional layers. 
At this stage we get \emph{segment-level} outputs, which are then combined through a class-specific attention mechanism to produce a \emph{recording-level} output.
The network is trained using a sequential self-teaching approach leading to robust generalization.
We use WEANet's $2,048$-dimensional hidden representation from the output of block B5 (average/max pooled over time and $l_2$-normalized) for transferring to target tasks. 



\section{Transfer Learning to Target Tasks} \label{sec:target-tasks}

Our motivation here is to understand the knowledge transfer from SED to a variety of other audio downstream tasks, focusing on sounds, actions, music, etc., and with small as well as large-scale datasets. The representations from TALNet and WEANet are used to train \textbf{linear classifiers} for these tasks. This helps us in focusing on representative power of the learned representations for the target tasks rather than relying on strong classifiers to obtain good performance.  

\subsection{Sound Event Classification}

Although SED on AudioSet is our source task, we also consider sound event classification on $3$~other datasets as target tasks: ESC-50~\cite{piczak2015esc}, Urbansound~\cite{salamon2014dataset} and \fsd~\cite{fonseca2019audio}. The domain mismatch between these datasets and AudioSet makes transfer learning non-trivial.
The \fsd dataset, in particular, is more challenging.
Unlike the other two, \fsd is a multi-label dataset, where each recording can have more than one label. 
It also consists of a ``curated'' set and a ``noisy label'' set: the former contains audio recordings carefully labeled by humans, whereas the latter can have wrongly labeled audio examples. An estimated $60\%$ of all the labels are wrong, making it a very challenging task.

\subsection{Acoustic Scene Classification}
Acoustic scenes are often composed of a mixture of sounds thereby exhibiting complex acoustic characteristics. While this implicit relation between sound events and acoustic scenes can provide some nuanced understanding of acoustic scenes, it remains to be seen if representations based on SED can capture enough information for good scene classification performance. Here we evaluate the transferability of SED to acoustic scene classification using Task~1a of the 2019 DCASE challenge~\cite{mesaros2018dcase}.

\subsection{Music Tagging}

This target task aims at tagging audio recordings with different music genres, instruments, moods, etc. It adds on the variability of target labels considered in this work. We use the well-known \magna dataset~\cite{law2009evaluation} for transfer evaluation. This is a multi-label dataset, where each recording can belong to a genre class as well as multiple instrument classes, at the same time. We use the top $50$~tags of this dataset in our experiments.

\subsection{Human Action Classification Using Audio}

The goal of this task is to recognize human actions such as ``ice skating'' and ``playing guitar'' in video recordings. We use the most recent version of the Kinetics dataset (Kinetics700), a widely used benchmark for action classification~\cite{kay2017kinetics}. It is a large-scale dataset with over $550$k $10$-second clips from $700$~action classes. This problem has been primarily tackled from a visual perspective, although some multimodal approaches have also been proposed~\cite{ghanem2018activitynet, xiao2020audiovisual}.

In this paper, we explore \emph{audio-only} recognition of human actions. This is interesting in several aspects. To the best of our knowledge, this is perhaps the \emph{first work} that explicitly tries to link human actions and sound events. In principle, similar to ImageNet~\cite{deng2009imagenet} based pre-trained models being used for visual-driven action classification, we hypothesize that pre-trained SED models can help advance the state of audio-driven action classification. 
Further, being a large-scale dataset with over $550$k clips, transferring SED representations to this task via linear classifiers helps characterize the efficacy of direct classification of actions vs. action classification based on knowledge of sounds. 


\begin{table}[t!]
  \centering
    \begin{tabular}{c|c|c}
    \hlineB{2}
	\textbf{Method} & \textbf{MAP} & \textbf{MAUC} \\
    \hline
	Ford {\em et. al}~\cite{ford2019deep} & 0.380 & 0.970 \\
	\hline
	TALNet~\cite{wang2019comparison} & 0.386   & 0.971 \\ 
    \hline
	WEANet-SUSTAIN~\cite{kumar2020sequential} & 0.398 & 0.972 \\
    \hlineB{2}
    \end{tabular}
    \caption{Comparison with state-of-the-art methods on AudioSet}
  \label{tab:AudioSet}%
  \vspace{-0.3in}
\end{table}%

\begin{table*}[t!]
\centering
\begin{tabular}{c|c|c|c|c|c|c|c|c}
\hlineB{2}
\textbf{Task} & \textbf{Dataset} & \textbf{\# Classes} & \multicolumn{2}{c|}{\textbf{Metric}} & \textbf{TALNet} & \textbf{WEANet} & \multicolumn{2}{c}{\textbf{Prior Work (Uses TL?)}} \\
\hline
\multirow{4}{*}{Sound Events} & ESC-50 & 50 & \multicolumn{2}{c|}{Accuracy} & 91.0 & 94.1 & 94.7~\cite{kong2019panns} & \cmark \\
\cline{2-9}
& Urbansound & 10 & \multicolumn{2}{c|}{Accuracy} & 85.2 & 85.2 & 85.1~\cite{palanisamy2020rethinking} & \cmark \\
\cline{2-9}
& \multirow{2}{*}{\fsd} & \multirow{2}{*}{80} & \multirow{2}{*}{lwlrap} & Curated & 72.0 & 72.8 & 54.2~\cite{fonseca2019audio} & \multirow{2}{*}{\xmark} \\
\cline{5-8}
& & & & Noisy & 51.0 & 50.3 & 31.2~\cite{fonseca2019audio} & \\
\hline
Acoustic Scenes & DCASE2019 & 10 & \multicolumn{2}{c|}{Accuracy} & 65.8 & 68.0 & 58.9~\cite{kong2019panns} & \cmark \\
\hline
Music Tagging & \magna & 50 & \multicolumn{2}{c|}{MAUC} & 91.5 & 91.5 & 90.2~\cite{lee2017multi} & \cmark \\
\hline
\multirow{2}{*}{Human Actions} & \multirow{2}{*}{Kinetics700} & \multirow{2}{*}{700} & \multirow{2}{*}{Accuracy} & Top-1 & 15.9 & 18.0 & 21.9~\cite{qiu2019trimmed} & \multirow{2}{*}{\xmark} \\
\cline{5-8}
& & & & Top-5 & 30.5 & 33.0 & 36.9~\cite{qiu2019trimmed} & \\
\hlineB{2}
\end{tabular}
\caption{Summary of the target tasks and the performance of TALNet and WEANet-SUSTAIN, compared with some previous works.}
\label{table:performance}
\vspace{-0.1in}
\end{table*}

\section{Experiments} \label{sec:exp}

\subsection{Datasets and Setup}

For fair comparison, we follow the standard training/validation/test split and use performance metrics defined for each dataset. When such information is unavailable, we follow the most prevailing setup from previous works. For ESC-50 and Urbansound, we perform $5$-fold and $10$-fold cross-validation following the predefined folds, and report the average accuracy across all folds. For \mbox{\fsd}, the ``public'' test set is used for validation. 
For \magna, we use the $12:1:3$ split for training, validation and testing, as was done in several prior works~\cite{dieleman2014end, wang2019music, wang2020hierarchical}. For Kinetics700, we take out $20,525$ examples from the training set to use as a validation set.
All models are implemented in PyTorch, and hyperparameters are tuned using the validation sets. 

\subsection{AudioSet Models}

The details of the TALNet and WEANet models trained on AudioSet can be found in~\cite{wang2019comparison} and \cite{kumar2020sequential} respectively. Table~\ref{tab:AudioSet} shows the performance of the two models on AudioSet. In this work, we re-trained TALNet applying SpecAugment~\cite{park2019specaugment} to the inputs. We masked out one frequency band of at most $16$~bins, and one time interval of at most $2$~seconds. This improves the the mean average precision (MAP) from $0.359$ in~\cite{wang2019comparison} to $0.386$. 

\subsection{Results}

Table~\ref{table:performance} summarizes results for all target tasks. For brevity, we also show performance from prior works that most closely relate to the proposed approach. To our knowledge, no such transfer learning work exists for \fsd and Kinetics700 (shown by \xmark under the ``Uses TL?'' column); for the other tasks, the reported baselines are from prior works using transfer learning. However, these transfer learning processes are often much more complex compared to our simple linear classifiers trained on learned representations from TALNet and WEANet. 

For the target tasks of sound event classification, the linear classifiers built upon TALNet and WEANet representations give similar performance as prior transfer learning works on the ESC-50 and Urbansound datasets. These numbers also come close to the state-of-the-art (SOTA) on these datasets. Note that, \cite{palanisamy2020rethinking} applies transfer via networks pre-trained on images. On \fsd, we compare with the baseline approach in \cite{fonseca2019audio}, and our results are $34\%$ and $63\%$ superior. 

For the target task of acoustic scene classification, audio representations from TALNet and WEANet give $6.9\%$ and $9.1\%$ better performance compared to \cite{kong2019panns}] (also trained on AudioSet). 

On the music tagging task, TALNet- and WEANet-based representations lead to better performance compared to the transfer learning proposal from~\cite{lee2017multi}. Interestingly, \cite{lee2017multi} uses a large-scale music tagging as source task (the Million Song Dataset \cite{bertin2011million}), which is very similar to the target task. While AudioSet also contains a fairly large number of music examples, this clearly shows that it is possible to construct good representations for music tagging via a more general-purpose source task like SED.

For Kinetics700, the performance of linear classifiers with audio representations from TALNet and WEANet is inferior to training an Xception model from scratch \cite{qiu2019trimmed}. This is expected for a large-scale dataset such as Kinetics700. But it is noteworthy that these audio representations can give competitive results with just linear classifiers, illustrating the shared information between the two tasks -- this has not been previously studied or observed. 
Some action classes, such as ``rolling eyes'' and ``peeling banana'', do not exhibit specific acoustic signatures and are hard to detect through any audio-only approach. The action ``playing bagpipes'' achieves the highest top-1 accuracy of $87.5\%$ (using WEANet features). This is not surprising because the ``bagpipes'' event gets the highest performance on the source AudioSet task as well. In Sec.~\ref{sec:correlation} we will provide some qualitative interpretation of the relationship between the source SED task and target Kinetics700.

\section{Analysis and Visualizations}
\label{sec:visualization}

Overall, the results on target tasks shows that, in most cases, simple linear classifiers that transfer TALNet and WEANet representations can give competitive, or marginally better, results compared to previously published numbers on these datasets. To bring further insights, we provide some more analysis here. 

We first show that the representations can capture semantics-driven proximity relationships among the target labels. We show this through the linear classification weights learned for each class in the target task. We also analyze the correlation between the target task labels and source task sound events, and illustrate that to a certain extent we can explain the sound events that contribute to specific target labels. 
Keeping page-limit in mind, we summarize the analysis for TALNet-based representations alone; similar results were obtained for WEANet representations.

\subsection{Clustering of Target Labels}
\label{sec:clustering}

The weight matrix of the linear model learned for a target task is essentially a {\it condensed} representation of the target labels' semantics. We denote this as $W \in \mathbb{R}^{C \times D}$, where $C$ is the number of classes in the target task and $D$ is the dimensionality of audio representations. 

\begin{figure}[!t]
  \centering
  \includegraphics[width=\columnwidth]{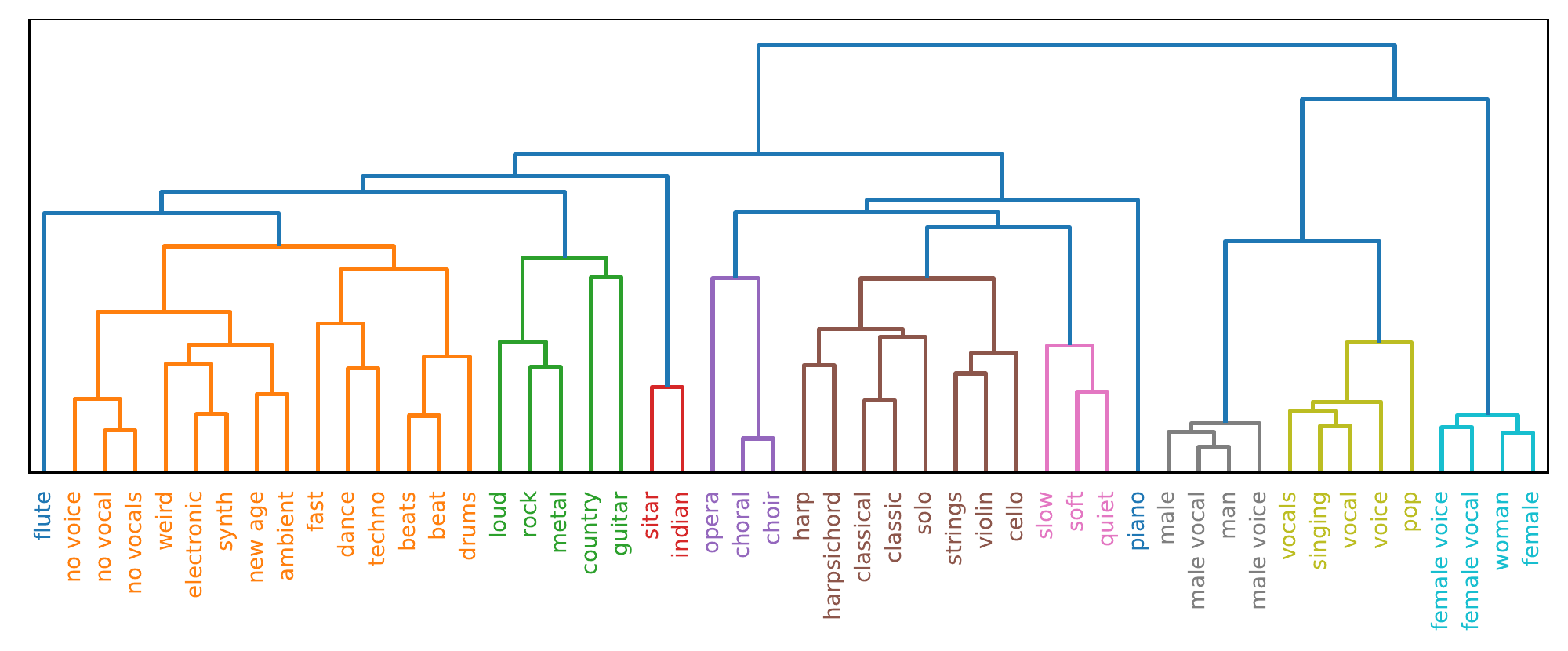}
  \vspace{-0.3in}
  \caption{Hierarchical clustering dendrogram of the \magna music tags.}
  \label{fig:magna-hier}
  \vspace{-0.1in}
\end{figure}

Consider the music tagging task with TALNet representations as an example. The learned weight matrix $W$ has a size of $50 \times 1,024$; each $1,024$-dimensional row vector essentially represents a music tag. If the TALNet representations do allow for learning the semantics of the music tags, then in this $1,024$-D space, semantically similar tags should be close to each other.
Given this hypothesis, we perform a hierarchical clustering in this space. The resultant dendrogram is shown in Fig.~\ref{fig:magna-hier}. It clearly shows the hypothesized semantically meaningful grouping of classes.
In particular, we see that synonymous tags such as ``woman'', ``female'', ``female vocal'', and ``female voice'' are clustered together. Similarly, instruments of classical music (\eg ``violin'' and ``harp'') form a cluster, and so do words describing vibrant music (\eg ``drums'', ``beat'', and ``dance''). This shows that our setup and source task are robust in learning general task-agnostic (abstract) information about audio and sounds. 

\begin{figure}[!t]
  \centering
  \includegraphics[width=\columnwidth]{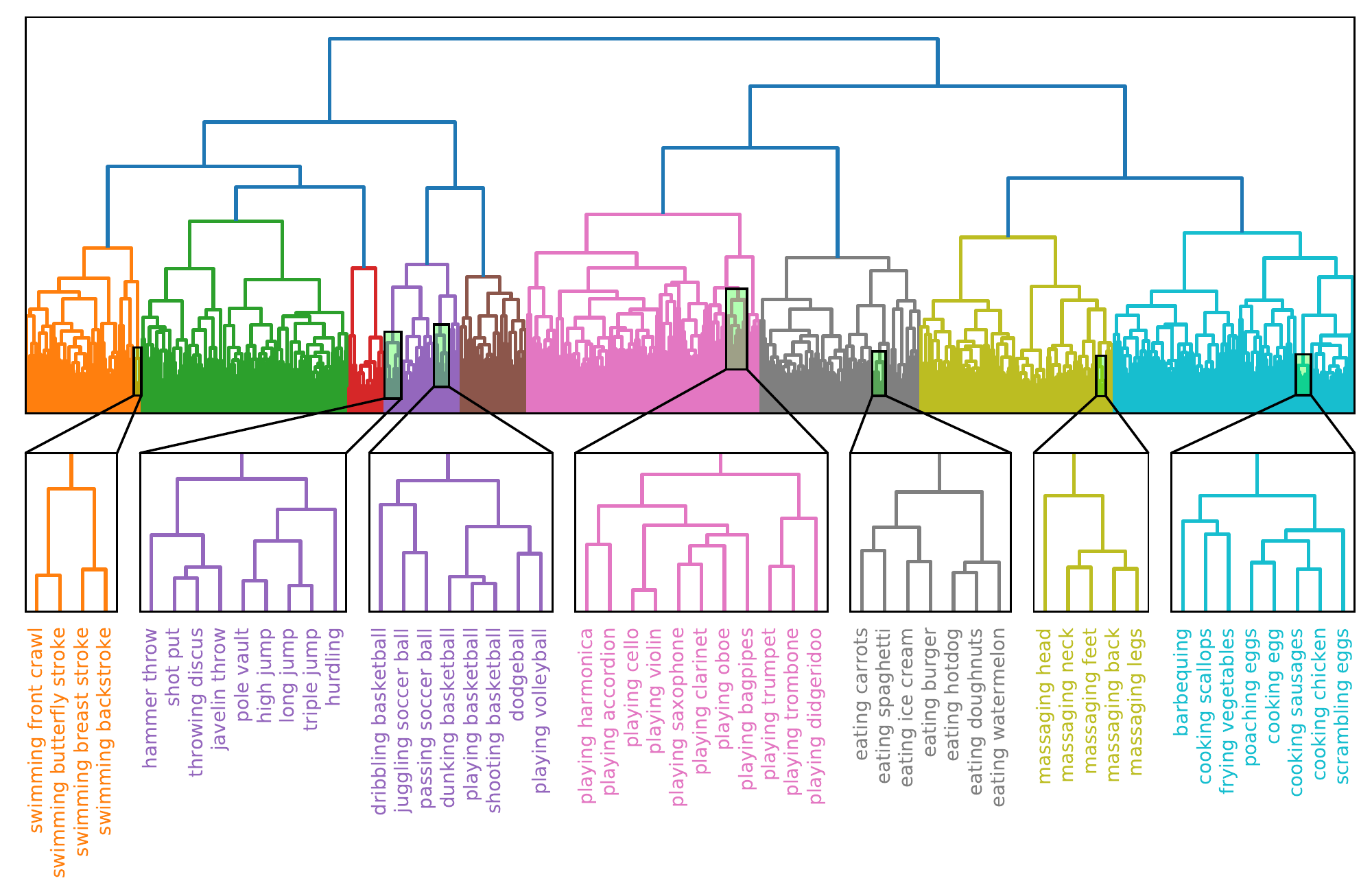}
  \vspace{-0.3in}
  \caption{Hierarchical clustering dendrogram of the Kinetics700 actions. To avoid clutter, action names are only shown for some small clusters.}
  \label{fig:kinetics-hier}
\end{figure}

\begin{figure}[!t]
  \centering
  \vspace{-0.2in}
  \includegraphics[width=0.75\columnwidth]{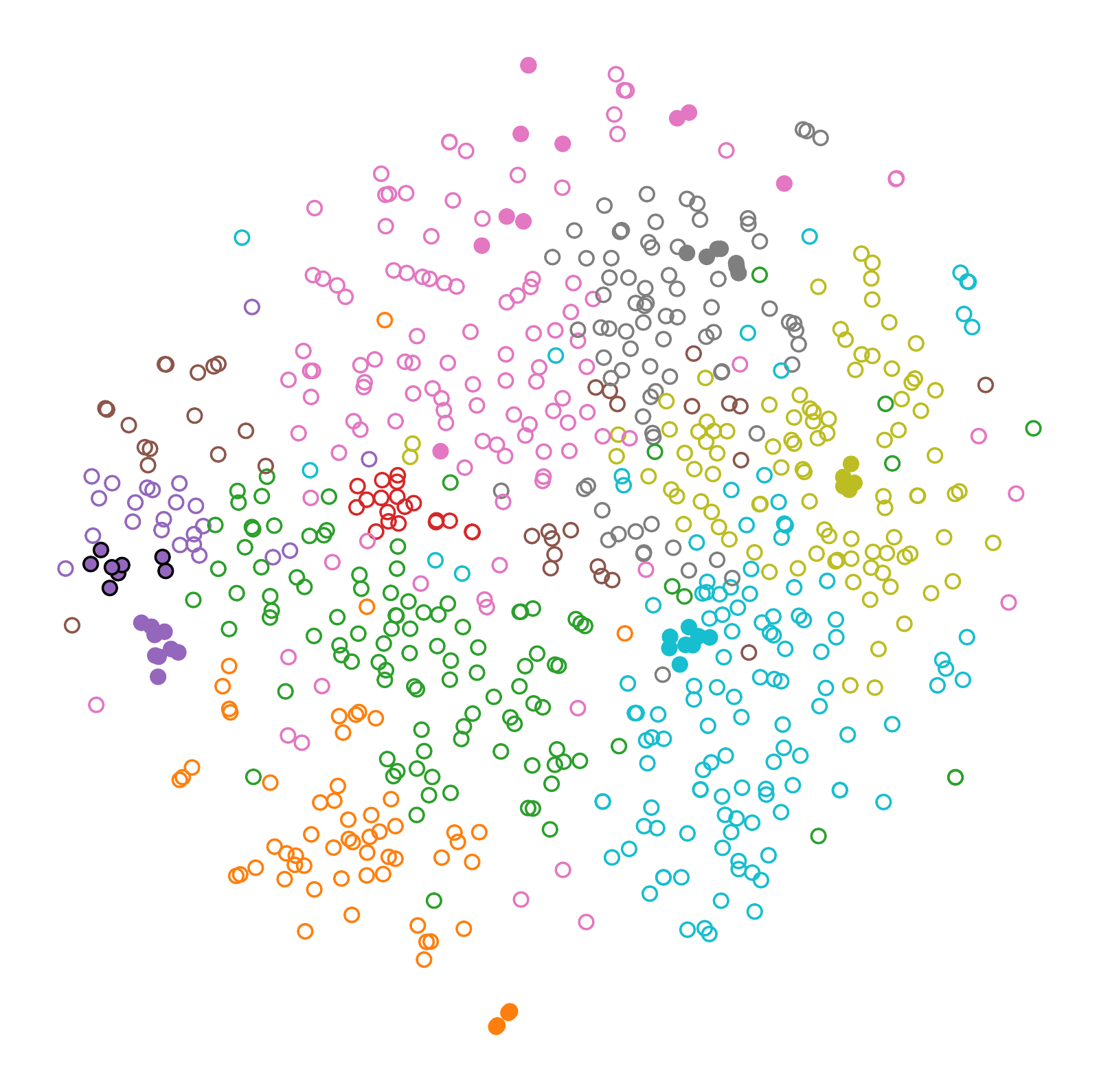}
  \vspace{-0.25in}
  \caption{t-SNE plot of the Kinetics action classes.}
  \label{fig:kinetics-tsne}
  \vspace{-0.15in}
\end{figure}

We performed a similar analysis for the human action recognition task, and the resulting dendrogram is shown in Fig.~\ref{fig:kinetics-hier}. Given the rather large number of action types in Kinetics700, we only show a few action names. 
Observe that we can recognize semantic clusters at both macro and micro levels. At the macro level, one can summarize each large cluster (marked by colors) with a few words. For example, most of the actions in teal are \emph{housework}, and most actions in purple are \emph{sports on land}. 
At a finer level of resolution, we can see small clusters representing \emph{ball sports} and \emph{track-and-field} sports form within the large purple cluster. Surprisingly, action classes such as massaging head, neck, back, legs, and feet, which do not correspond to sound signals, are also well clustered. A possible explanation is that these actions often come with audio tracks containing relaxing music, and the audio representations are able to exploit such acoustic cues to support the recognition of visual actions. Fig.~\ref{fig:kinetics-tsne} provides an alternate view by running t-SNE~\cite{maaten2008visualizing} on the learned class representations. The color coding follows the dendogram coloring scheme. Once again, we notice that closely related events cluster together. In summary, deep audio representations from large scale SED task may directly be used to learn semantic relationships among human-actions using a linear classification transfer methodology. 

\subsection{Correlation Between Target Labels and Sound Events}
\label{sec:correlation}

While we have shown that the audio representations from our source task contain adequate information for recognizing music tags and actions, it is hard to interpret and rationalize the evidence that the transfer learning models use to predict a target label.
To understand this, we study the correlation between the target labels and sound events to see if predictions of target labels are often supported by the existence of certain sound events.

We compute the cosine similarity between the following two sets of vectors: $1,024$-D representations of music tags and actions, taken from the rows of the weight matrices of the two linear classifiers; and the $1,024$-D representations of the $527$ AudioSet sound events, taken from the rows of the weight matrix of the final fully connected layer of TALNet. Before computing the cosine similarity, we perform mean-variance normalization.

\begin{figure}[!t]
  \centering
  \includegraphics[width=\columnwidth]{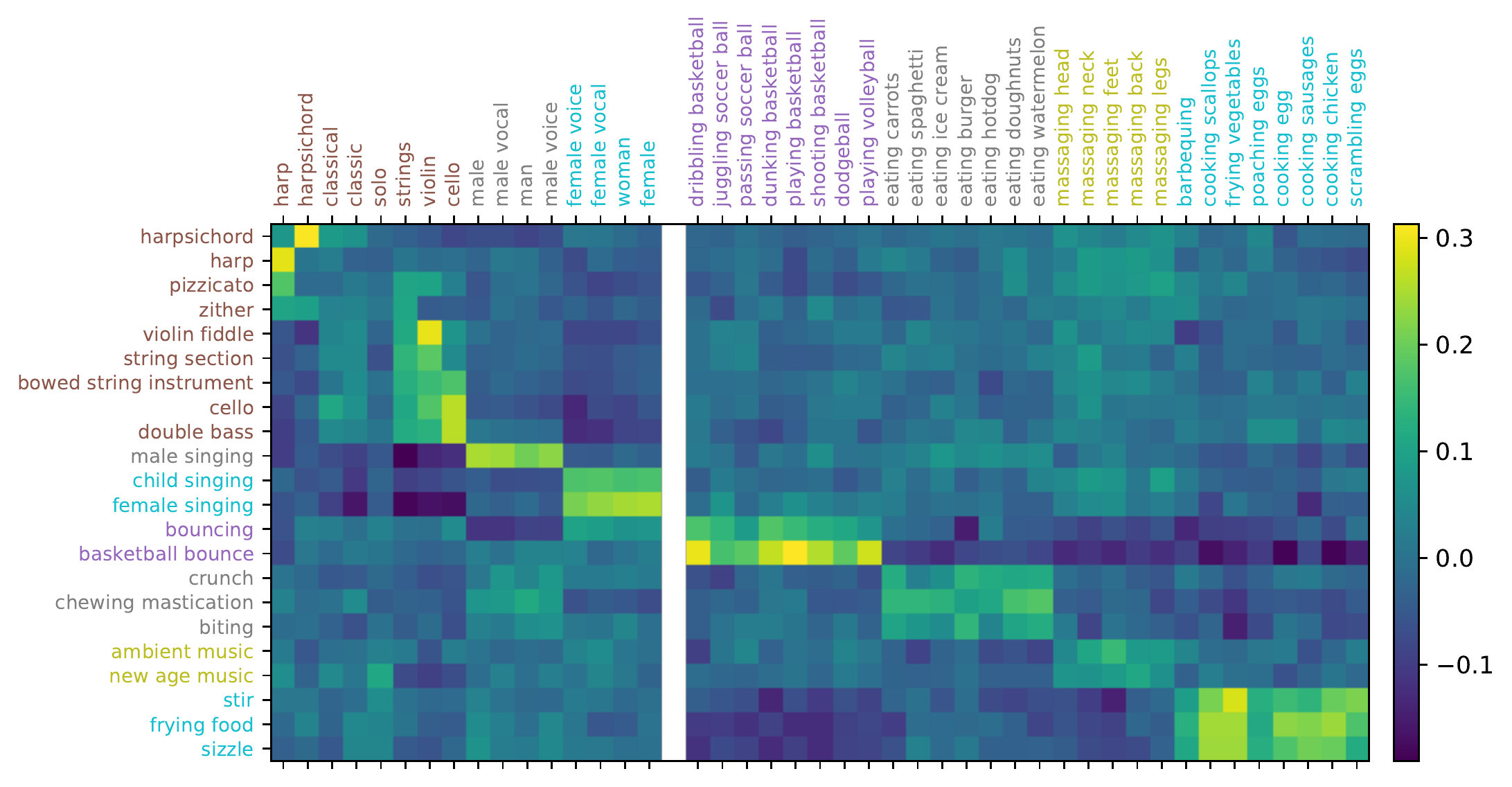}
  \vspace{-0.3in}
  \caption{Cosine similarity between some AudioSet sound events (rows) and \magna music tags / Kinetics700 actions (columns).}
  \label{fig:cosine}
  \vspace{-0.15in}
\end{figure}

A part of the resulting cosine similarity matrix is shown in Fig.~\ref{fig:cosine}. Rows represent sound events, and columns represent music tags and actions. The rows and columns are sorted according to the dendrograms produced by hierarchical clustering, so that similar music tags, actions, and sound events are next to each other. We can immediately recognize blocks of high similarity values (often $\ge 0.2$), manifesting themselves in yellow and light green cells. Considering that two random vectors in a high-dimensional space are usually nearly orthogonal, cosine similarity values above 0.2 are remarkably large. This figure demonstrates that many music tags or actions can be explained by a single or a few sound events, and transfer learning is able to discover such correspondences. For example, various actions of cooking exhibit high similarity to events like ``sizzle''; actions of eating are characterized by ``chewing''. Actions like massaging are often accompanied by relaxing music such as ``new age music'', although the similarity is not as high. 
Overall, the correlation analysis provides a quantifiable way to interpret the correspondences between music tags, actions, and sound events.  

\section{Conclusion}
\label{sec:conclusion}

We demonstrated that it is possible to transfer knowledge from sound event detection (SED) task to a wide range of other audio tasks, including acoustic scene classification, music tagging and human action recognition. 
Using a simple linear classifier on audio representations obtained from SED models, we are able to achieve performance that is comparable or better than the state of the art on several datasets. The linear classification system also provides a lucid way to interpret the suitability of these representations for target tasks. By visualizing the classifier learned representations learned for target tasks, we found meaningful structures that reflect proximity relationships among music genres, instruments, moods, as well as various human actions. 
Lastly, it was possible to identify the unique sound events that contribute to the learning of downstream tasks.

\vfill\pagebreak
\bibliographystyle{IEEEtran}
\bibliography{refs}


\end{document}